\documentclass[referee]{chjaa}
\usepackage{amsmath}          

\usepackage{graphicx,times}             
\input{epsf.sty}                        
\input{psfig.sty}                       

\begin{document}

   \title{Modeling $\epsilon$ Eridani and asteroseismic tests of element
   diffusion}


   \volnopage{Vol.0 (200x) No.0, 000--000}      
   \setcounter{page}{1}           

   \author{Ning Gai
             \inst{1,3}
   \and Shao-Lan Bi
        \inst{2,1}
   \and Yan-Ke Tang
         \inst{1,3}
      }

   \institute{National Astronomical Observatories/Yunnan Observatory, Chinese Academy of Sciences, Kunming 650011,
              P. R. China\\
             \email{ bisl@bnu.edu.cn; gaining@ynao.ac.cn}
        \and
             Department of Astronomy Beijing Normal University, Beijing 100875,
             P. R. China\\
        \and
             Graduate School of The Chinese Academy of Sciences, Beijing 100039, P. R. China\\
          }

   \date{Received~~2001 month day; accepted~~2001~~month day}

\abstract{Taking into account the helium and metal diffusion, we
explore the possible evolutionary status and perform seismic
analysis of MOST target: the star $\epsilon$ Eridani. We adopt the
different input parameters to construct the models by fitting the
available observational constraints: e.g., $T_{eff}$, $L$, $R$,
$[Fe/H]$. From computation, we obtain the average large spacings of
$\epsilon$ Eridani about $194\pm 1 \mu$Hz. The age of the diffused
models has been found to be about 1 Gyr, which is younger than one
determined previously by models without diffusion. We found that the
effect of pure helium diffusion on the internal structure of the
young low-mass star is slight, but the metal diffusion influence is
obvious. The metal diffusion leads the models to have much higher
temperature in the radiation interior, correspondingly the higher
sound speed in the interior of the model, thereby the larger
frequency and spacings.
   \keywords{stars: evolution --- stars: interiors
  --- stars: individual: $\epsilon$ Eridani}
   }

   \authorrunning{Gai, N. et al.}            
   \titlerunning{Modeling $\epsilon$ Eridani and asteroseismic tests of element
   diffusion }  

   \maketitle

%
%
\section{Introduction}           
\label{sect:intro}

Element diffusion, sometimes named ``atomic" diffusion, is a basic
physical element transport mechanism which is driven by pressure
gradients (or gravity), temperature gradients and composition
gradients. The effects of helium diffusion on the structure of solar
models were first studied by Noerdlinger (\cite{Noerdlinger}). Later
helioseismology provided an effective method to study this physical
process (Guenther et al. \cite{Guenther93}; Guenther
\cite{Guenther94}; Bahcall \& Pinsonneault \cite{Bahcall92b};
Christensen-Dalsgaard et al. \cite{Christensen-Dalsgaard93};
Guenther \& Demarque \cite{Guenther97}). Recently, helioseismic
inferences have demonstrated that models incorporating element
diffusion are in substantially better agreement with the inferred
sound speed than the models that neglect it.

Up to date, the effect of element diffusion has been tested by
asteroseismic method in solar type stars. According to the ``second
differences", Vauclair \& Th\'{e}ado (\cite{Vauclair}), Th\'{e}ado
et al. (\cite{Theado}) and Castro \& Vauclair (\cite{Castro})
presented discussion of asteroseismic signatures of pure helium
diffusion in main-sequence stars between 1.1$M_{\odot}$ and
2.0$M_{\odot}$. In the present work, we study the effects of both
helium and metal element diffusion on the models with mass less than
1$M_{\odot}$.

Metal element could affect temperature gradient and convection in
the envelope of stars. It also leads to the variations of the
central temperature, density, pressure and radiative opacity. It is
important to consider the influence of metal element on the internal
structure of star. In this paper we mainly study the effects of
metal diffusion on the stellar structure, evolution and the
oscillation frequency for the late K-type star $\epsilon$ Eridani.

The bright K2V dwarf $\epsilon$ Eridani (HD22049, HIP16537, HR1084,
PLX742) is one of the nearest solar-like stars, with a distance of
about 3.218pc. It has a planetary companion $\epsilon$ Eridani b
(Hatzes et al. \cite{Hatzes}). Its proximity makes this planetary
system as a prime target for future extrasolar planet direct --
imaging efforts and studying the process of planetary systems
formation. The success of these efforts will depend on the mass and
the age of the system. Therefore the study of exoplanet -- host star
$\epsilon$ Eridani is a key issue.

In order to determine precise global parameters of the star
$\epsilon$ Eridani, some authors performed numerous theoretical
analysis (Guenther \& Demarque \cite{Guenther86}; Guenther
\cite{Guenther87}; Soderblom \& D\"{a}ppen \cite{Soderblom})
(hereafter GD86, G87 and SD89 respectively). Thanked for  Noyes et
al. (\cite{Noyes}), they measured the $p$-mode oscillation spectrum
of $\epsilon$ Eridani and identified three individual frequencies in
the power spectrum between 1500 $\mu$Hz and 2000 $\mu$Hz with
average large spacing of 172 $\pm$ 5 $\mu$Hz. Based on
non-asteroseismic and asteroseismic observational constrains, GD86,
G87 and SD89 constructed a series of stellar models of $\epsilon$
Eridani. However, in their works for $\epsilon$ Eridani, there is a
contradiction between the young age implied by the high
chromospheric activity, rapid rotation rate, and the old age implied
by the slightly metal-poor composition. In their works, GD86, G87
and SD89 discussed this contradiction and given the age 12 Gyr, 10
Gyr and 1 Gyr respectively.

Recently many evidences from observations seem to indicate that
$\epsilon$ Eridani is a young main-sequence star with age less than
1 Gyr. For example, using Li abundances with the star's position in
the H-R diagram and kinematics, Song et al. (\cite{Song}) derived
the age of 0.73 $\pm$ 0.2 Gyr; Measuring the radius of $\epsilon$
Eridani by long-baseline interferometry, Di Folco et al. (\cite{Di
Folco}) estimated the age at 0.85 Gyr; Saffe et al. (\cite{Saffe})
used the calibrations of Donahue (\cite{Donahue93}) and Rocha-Pinto
\& Maciel (\cite{Rocha-Pinto}) (which corrected the age with an
effect from chromospheric activity) to estimate the ages of 0.66 Gyr
and 0.82 Gyr respectively. SIMBAD database describes the age of
$\epsilon$ Eridani to be 0.66 Gyr.

The aim of present paper is to provide the basic theoretical studies
of $\epsilon$ Eridani. We will use the latest interferometric
observational radius (Di Folco et al. \cite{Di Folco}) to constraint
the stellar models and predict the frequency spacings. Then, we will
investigate the effects of helium and metal diffusion on the
internal structure, and the frequency spacings which also depend on
the internal structure of the star.

In section 2, the global parameters of the star were summarized.
The input physics, computing method and results analysis are
summarized in section 3. The helium and metal diffusion effects on
the stellar models, and the oscillation frequency are discussed in
section 4. Finally, the conclusion is presented in section 5.

\section{Observational Constraints}

Observationally it is difficult to determine accurately the mass of
star except some binary systems. $\epsilon$ Eridani is a single star
and it has various data of masses which were published in the
literatures, for example 0.85 $\pm$ 0.04$M_{\odot}$ (Santos et al.
\cite{Santos01}), 0.80$M_{\odot}$ (Minier \& Lineweaver
\cite{Minier}), 0.82$M_{\odot}$ (Takeda et al. \cite{Takeda}),
0.83$M_{\odot}$ (Fischer \& Valenti \cite{Fischer}).

For the effective temperature of $\epsilon$ Eridani, we adopt
$T_{eff}$ = 5012 $\pm $ 67 which was determined from the infrared
flux method (IRFM) (Ram\'{\i}rez \& Mel\'{e}ndez \cite{Ramirez}).

The luminosity can be calculated using the absolute visual magnitude
$M_{v}$ = 6.18 $\pm$ 0.11 (Song et al. \cite{Song}; Takeda et al.
\cite{Takeda}). Given the bolometric correction $B.C.$ = -- 0.28
(Takeda et al. \cite{Takeda}), the absolute bolometric magnitude is
therefore $M_{bol}$ = $M_{v}$ + $B.C.$ = 5.9 $\pm$ 0.11. Then the
luminosity is computed by formulae: $M_{bol}$ = --
2.5$log(L/L_{\odot})$ + $M_{bol, \odot}$. Adopting $M_{bol, \odot}$
= 4.746, we thus obtain $\log$ ($L/L_{\odot}$) = -- 0.4176 $\sim$ --
0.5056.

$\epsilon$ Eridani is slightly metal-poor compared with the Sun.
Santos et al. (\cite{Santos04}) obtained $[Fe/H]$ = -- 0.13 $\pm$
0.04 from the spectroscopic analysis basing on 39 Fe$_{\rm I}$ and
12 Fe$_{\rm II}$ lines. In order to deduce $[Z/X]$, we assume that
the logarithmic value of $[Z/X]$ is proportional to the abundance
ratio $[Fe/H]$ (Thoul et al. \cite{Thoul03}):
\begin{equation}
\log[Z/X]_{star} = [Fe/H]_{star} + \log[Z/X]_{\odot}
\end{equation}
where $[Z/X]_{\odot}$ = 0.023 (Grevesse \& Sauval \cite{Grevesse}).
Using this ratio, we obtain $[Z/X]_{surf}$ = 0.017 $\pm$ 0.0016.

A star's initial helium mass fraction $Y_{i}$ is important for
determining its structure, but $Y_{i}$ of $\epsilon$ Eridani is
unknown by observations. GD86, G87 adopted $Y_{i}$ to be 0.236 and
SD89 considered $Y_{i}$ to be from 0.24 to 0.26. In our work, we
reference the value given by GD86, G87, SD89 and adopt the initial
helium abundance $Y_{i}$ = 0.245 $\pm$ 0.015.

The frequency, especially large spacings, sensitively depend on the
radius. In order to predict the large spacings which will be
observed by MOST, we take the latest interferometric observational
radius $R/R_{\odot}$ = 0.734 $\pm$ 0.01 (Di Folco et al. \cite{Di
Folco}).

\begin{table}
 \caption{The  observational fundamental parameters of
$\epsilon$ Eridani.
            }
\label{table : 1}
 \tabcolsep0.15in
\begin{tabular}{c| c c}
\hline\hline
 parameters &$\epsilon$ Eridani & Ref \\
 \hline

 $M/M_{\odot}$&$0.85 \pm 0.04$&(1)\\
                   &$0.80$&(2)\\
                   &$0.82$&(3)\\
                   &$0.83$&(4)\\
\\

 $[F_{e}/H]_{surf}$&$-0.13 \pm 0.04$&(5)\\
\\

 $R/R_{\odot}$&$0.743 \pm 0.010$&(6)\\

 $M_{v}$&$6.18 \pm 0.11$&(7)\\
                  &$6.18$&(3)\\

 $\log(L/L_{\odot})$&$-0.4616 \pm 0.044$&(8)\\

 $T_{eff}(K)$&$5012 \pm 67 $&(9)\\

 \hline\hline

 \end{tabular}

 (1)Santos et al. (\cite{Santos01});
 (2)Minier \& Lineweaver (\cite{Minier});
 (3)Takeda et al. (\cite{Takeda});
 (4)Fischer \& Valenti (\cite{Fischer});
 (5)Santos et al. (\cite{Santos04});
 (6)Di Folco et al. (\cite{Di Folco});
 (7)Song et al. (\cite{Song});
 (8)this paper;
 (9)Ram\'{\i}rez \& Mel\'{e}ndez (\cite{Ramirez})

 \end{table}

\section{Stellar modeling}
\subsection{Input physics}

The evolutionary models are computed using the Yale stellar
evolution code version (Guenther et al. \cite{Guenther94}) which has
been modified to include the effects of element diffusion. The
initial zero-age main sequence (ZAMS) models were calculated from
pre-main sequence evolution which were assumed to have spherical
symmetry, no rotation, no magnetic field. In the computation we use
OPAL equation of state tables EOS2001 (Rogers \& Nayfonov
\cite{Rogers}), the opacities with a smooth blend of OPAL GN93
(Iglesias \& Rogers \cite{Iglesias}) and OPAL tables (Alexander \&
Ferguson \cite{Alexander}). The relevant nuclear reaction rates and
cross sections are from Bahcall \& Pinsonneault (\cite{Bahcall92a},
\cite{Bahcall92b}) and Bahcall (\cite{Bahcall89}). $\epsilon$
Eridani is a solar-like star, we choose the Krishna-Swamy
(\cite{Krishna Swamy}) atmosphere model. The diffusion of both
helium and metal element abundance was considered in the stellar
model computation, by using the coefficients of Thoul et al.
(\cite{Thoul94}). In the next section we will introduce it in
detail.

 Using the standard mixing-length theory, we
set $\alpha$ = 1.7 for all models, close to the value which is
required to reproduce the solar radius under the same physical
assumptions and stellar evolution code (Murphy \& Demarque
\cite{Murphy}). If rotation and magnetic field are neglected in the
construction of stellar model, convection is the most important
mechanism that is usually included in standard stellar models which
describes material motions within a star especially for low-mass
main sequence star. The convective overshoot is an another factor to
influence the chemical composition of the star. The change of
mixing-length parameter $\alpha$ would also influence principally
the radius of the base of the external convection zone. The presence
of core overshoot would extend the core-burning phase of evolution
and increase the estimated age of star. This important topic has
been studied separately by Bi et al. (\cite{Bi}). However, the
element diffusion could also lead to the redistribution of the
element in star and change the depth of the convective envelope
(Guenther \cite{Guenther94}). In order to test the effect of
diffusion on the chemical composition, internal structure and the
frequency of $\epsilon$ Eridani, we would not discuss the influence
of the convection and overshoot in present work.

\subsection{Element diffusion}

The element diffusion in stars is driven by gravitational setting,
temperature gradients, composition gradients and radiation pressure.
It is described by Bahcall \& Pinsonneault (\cite{Bahcall92a}) in
detail. The gravity and temperature gradients tend to concentrate
the helium and heavier elements toward the center of the star, while
the hydrogen diffuses outward. The concentration gradients oppose
the above processes.  In addition, radiation pressure can cause
partially ionized or neutral species to rise relative to species
with a small cross section. However, the radiation pressure are
efficient in the external regions of main-sequence stars with
$T_{eff}>6000K$ (Michaud \cite{Michaud}). So the radiative pressure
causes negligible diffusion in the K-type star and will be neglected
in this paper.

\subsubsection{Basic equation for element diffusion}

Most works on chemical diffusion use either the Chapman - Enskog
procedure (Chapman \& Cowling \cite{Chapman}) or the method of
Burgers (\cite{Burgers}) for deriving the transport properties from
the Boltzmann equation. In our work the element diffusion is
described using the equations of Burgers (\cite{Burgers}) with
diffusion velocity coefficients from Thoul et al (\cite{Thoul94}). A
single model includes both helium and metal element diffusion. The
change rate of the element mass fractions due to the diffusion is
now written as (Thoul et al \cite{Thoul94}; Bahcall \& Loeb
\cite{Bahcall90}):
\begin{equation}
 \frac{\partial X_{s}}{\partial t} = -
\frac{1}{\rho r^{2}} \frac{\partial
[r^{2}X_{s}T^{5/2}\xi_{s}(r)]}{\partial r}
\end{equation}
where the partial derivatives are evaluated in the local rest frame
of a mass shell in the star. $\rho$, $r$, $T$ are the local value of
density, radius, temperature and $X_{s}$ is the mass fraction of
element $s$.

The diffusion velocity for species $s$ is defined by

\begin{equation}
w_{s} = \frac{T^{5/2}\xi_{s}}{\rho}
\end{equation}
and the function $\xi_{s}(r)$ is expressed as
\begin{equation}
\xi_{s}(r) = A_{p}(s)\frac{\partial\ln p}{\partial r} +
A_{T}(s)\frac{\partial\ln T}{\partial r} + \sum\limits_{c\neq e,2}
A_{c}(s)\frac{\partial\ln C_{c}}{\partial r}
\end{equation}
In equation (4), $A_{p}$, $A_{T}$, $A_{c}$ correspond to
gravitational setting, thermal diffusion and concentration gradient
diffusion which are the functions of the mass fractions only.
Species $c$ can equal to $s$, but $c$ aren't electron and helium
element.

 \subsubsection{The treatment of diffusion in the evolutionary
code}

 The Yale stellar evolution code has been modified to include
the effects of $He$ diffusion from gravitational settling and
thermal diffusion (Bahcall \& Pinsonneault \cite{Bahcall92a}) using
the method of Bahcall \& Loeb (\cite{Bahcall90}) and added the metal
diffusion in 1994. About the method by which treatment of element
diffusion in a standard stellar evolution code, it has been
described by Bahcall \& Pinsonneault (\cite{Bahcall92a}) in detail.
In this section, we summarize it again.

In the diffusion subroutine written by Bahcall and Pinsonneault, the
main problem is to solve the diffusion equation and calculate the
change of the element abundance by diffusion. The diffusion equation
(2) can be written as following form:
\begin{equation}
\frac{dx}{dt}=\frac{1}{\rho
r^{2}}[\frac{d}{dr}(D_{1})+\frac{d}{dr}(D_{2}\frac{dx}{dr})]
\end{equation}
where $D_{1}$, $D_{2}$ is the diffusion coefficients.

For helium diffusion

\begin{equation} \label{eq:1}
\left\{ \begin{array}{l}
         D_{1}(Y)=\frac{F_{gy}\cdot r^{2}\cdot T^{5/2}}{\ln\Lambda}\cdot\frac{d\ln p}{dr}\cdot X\cdot(A_{p}^{X}+A_{T}^{X}) \\
                  D_{2}(Y)=\frac{F_{gy}\cdot r^{2}\cdot T^{5/2}}{\ln\Lambda}\cdot A_{c}^{X}
                          \end{array} \right.
                          \end{equation}

For metal diffusion
\begin{equation} \label{eq:1}
\left\{ \begin{array}{l}
         D_{1}(Z)=\frac{F_{gz}\cdot r^{2}\cdot T^{5/2}}{\ln\Lambda}\cdot\frac{d\ln p}{dr}\cdot Z\cdot(A_{p}^{Z}+A_{T}^{Z}) \\
                  D_{2}(Z)=\frac{F_{gz}\cdot r^{2}\cdot T^{5/2}}{\ln\Lambda}\cdot A_{c}^{Z}
                          \end{array} \right.
                          \end{equation}
and radius, temperature, and density are in nondimensional units
defined by $r=r'/R_{\odot}$, $T=T'/10^{7}K$, $ \rho=\rho'/100g$
$cm^{-3}$ and $t=t'/10^{13}yr$ (Bahcall \& Pinsonneault
\cite{Bahcall92a}; Chaboyer et al. \cite{Chaboyer}).

In expressions (6) and (7), $F_{gy}$ and $F_{gz}$ are the adjustment
factors by which the helium and metal diffusion coefficients are
multiplied. In our work, we set both of the two factors to 1.0.
 $A_{p}$, $A_{T}$, $A_{c}$ are the diffusion velocity coefficients as described in equation (4).
 The exact
numerical solution of these coefficients are calculated by an export
subroutine which was developed by Thoul in 1994. The computation
method is exhibited detailedly by Thoul et al. (\cite{Thoul94}).

In equation (5) the partial time derivative is evaluated at constant
mass shell of the star. The diffusion equation is solved with zero
hydrodynamic velocity of the stellar plasma. The right hand of
diffusion equation (5) has two terms: the first term depends on
first spatial derivatives of the element mass fraction and the
second term depends on second spatial derivatives of the element
abundance. The first term is solved explicitly using the two-step
Lax-Wendroff technique (Press et al. \cite{Press}) by neglecting the
second derivatives of the element abundance. We then use this trial
solution as the initial abundance to determine the second
derivatives by a fully implicit method (Bahcall \& Pinsonneault
\cite{Bahcall92a}). At last, the element abundance were updated.

For the helium diffusion, we only consider $^{4}$He diffusion and
ignore $^{3}$He diffusion (Loeb et al. \cite{Loeb}).  The time rate
of change of the $^{4}$He mass fraction $Y$ is equal in magnitude
and opposite in sign to the rate of change of the hydrogen mass
fraction.

The diffusion of all metal elements were assumed to diffuse at the
same rate as fully ionized iron. Because of the diffusion of metal
elements, the radiative opacity have to be changed at each spherical
shell in the stellar model after each time step. We calculate the
effect of metal element diffusion on the opacity by computing a
total metal element abundance at each model radius $Z(r)$, and then
interpolating for the opacity between opacity tables with different
total metal element abundances (Bahcall et al. \cite{Bahcall95}).

The diffusion subroutine carries out the diffusion calculations
using data supplied by other parts of the Yale code, in which the
thermal structure and the element abundances are calculated. It is
assumed that the amount of diffusion within a given time step was
too small to affect significantly the changes in thermal structure,
in abundances, and in nuclear reaction rates that are calculated
elsewhere. After the diffusion computation, we update the element
abundance and then calculate the thermal structure.

In the diffusion subroutine, the element diffusion is treated only
in the radiation region. The characteristic time for element to
diffuse a solar radius under solar conditions is of the order of
$10^{13}$ yr (Bahcall et al. \cite{Bahcall95}), much larger than the
age of the sun and $\epsilon$ Eridani. However, helioseismic
inferences have demonstrated the significance of the element
diffusion (Cox et al. \cite{Cox}; Bahcall \& Loeb \cite{Bahcall90};
Bahcall \& Pinsonneault \cite{Bahcall92a}; Proffitt
\cite{Proffitt}), we would test how much effect on the internal
structure, the age and the frequency of $\epsilon$ Eridani by both
the helium and metal diffusion using the asteroseismic method.

\subsection{Computational method and results analysis }

\begin{table*}
 \caption{Model Parameters of
$\epsilon$ Eridani.
            }
\label{table : 1}
 \tabcolsep0.10in
\begin{tabular}{c| c c c|ccc|ccc}
\hline\hline

Model&A1&A2&A3&B1&B2&B3&C1&C2&C3\\
\hline

     &\multicolumn{3}{|c|}{$M/M_{\odot}=0.83$}&\multicolumn{3}{|c|}{$M/M_{\odot}=0.83$}&\multicolumn{3}{|c}{$M/M_{\odot}=0.83$}\\
     &\multicolumn{3}{|c|}{$Z_{i}=0.012$}&\multicolumn{3}{|c|}{$Z_{i}=0.013$}&\multicolumn{3}{|c}{$Z_{i}=0.014$}\\
     &\multicolumn{3}{|c|}{$Y_{i}=0.2315$}&\multicolumn{3}{|c|}{$Y_{i}=0.2365$}&\multicolumn{3}{|c}{$Y_{i}=0.2415$}\\
\hline

Dif.&None.&Y.Dif&Y\& Z.Dif&None.&Y.Dif&Y\& Z.Dif&None.&Y.Dif&Y\& Z.Dif\\
Age(Gyr)&1.151&1.089&1.0&1.206&1.138&1.0&1.264&1.191&1.0\\
$L/L_{\odot}$&0.321&0.321&0.321&0.319&0.319&0.319&0.317&0.317&0.317\\
$R/R_{\odot}$&0.749&0.750&0.744&0.750&0.751&0.743&0.751&0.753&0.741\\
$T_{eff}$&5024.9&5019.8&5039.5&5012.2&5006.7&5036.3&4999.9&4993.8&5033.5\\
&&&&&&&&&\\
$r_{cz}/R$&0.7073&0.7053&0.7048&0.7046&0.7033&0.7024&0.7024&0.7022&0.7000\\
$\tau_{cz}(s)$&1504.8&1514.1&1496.7&1514.4&1522.9&1496.7&1523.2&1529.5&1497.0\\
$\tau_{0}(s)$ &2524.5&2532.5&2502.4&2530.1&2538.8&2493.6&2535.7&2545.2&2484.5\\
$\tau_{cz}(s)/\tau_{0}(s)$&0.5961&0.5979&0.5981&0.5986&0.5999&0.6002&0.6007&0.6009&0.6025\\
$<\Delta \nu>$&192.2&191.4&193.8&191.6&191.0&194.6&191.2&190.5&195.1\\

 \hline\hline

 \end{tabular}

 \end{table*}

\subsubsection{Evolutionary tracks and candidates for pulsation}

\label{sect:intro}
 \begin{figure}
  \centering
  \includegraphics[angle=0,totalheight=5cm,width=8cm]{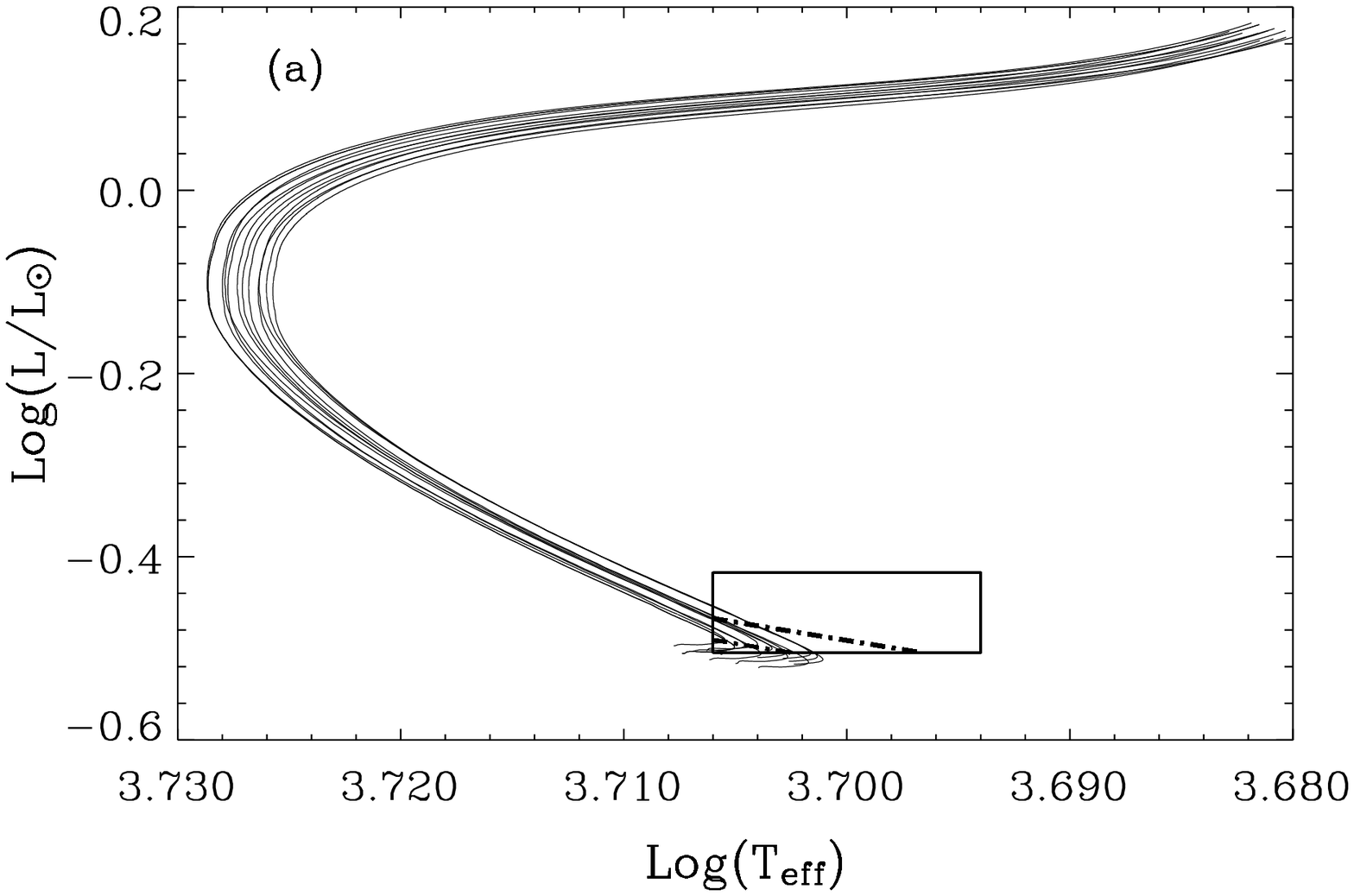}
  \vskip 0.15in
\includegraphics[angle=0,totalheight=5cm,width=8cm]{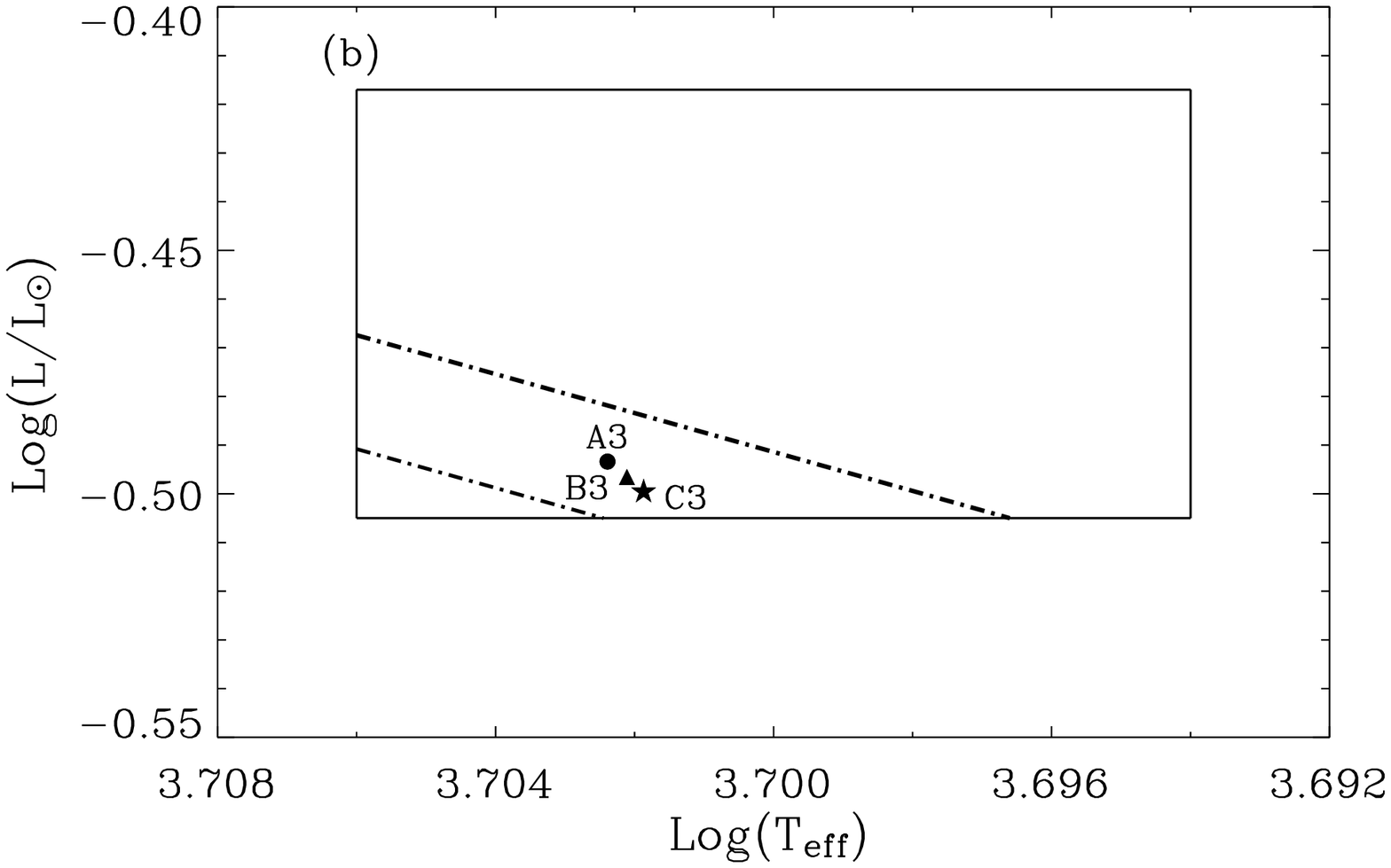}
\vskip 0.15in
  \caption{(a)Evolutionary tracks (12 in total) in the H-R diagram from ZAMS.
 The solid lines indicate the boxes delimited by the observed luminosity and effective temperature. The two dash dot lines denote the boxes delimited by the
 interferometric radius. (b)Enlarge the error box of Fig (a). The model A3, B3 and C3 are denoted by filled circle, triangle
 and five pointed star respectively.}%
   \end{figure}

In order to reproduce observational constraints of $\epsilon$
Eridani, we have computed a grid of evolutionary tracks for six
masses $M/M_{\odot}$: 0.80, 0.81, 0.82, 0.83, 0.84, 0.85 and the
initial metallicity $Z_{i}$: 0.011, 0.012, 0.013, 0.014.

From a series of evolutionary tracks which are computed, we select
those models that land within the observational error box (being
composed by $\log T_{eff}$, $\log L/L_{\odot}$, $\log R/R_{\odot}$)
in the theoretical H-R diagram. Then we further choose the models on
the basis of the constraint of the observational $[Z/X]$. In terms
of the computation alone, we note that the age of $\epsilon$ Eridani
is likely in the range ~ 0.1 Gyr -- 1.0 Gyr. Here we consider the
age ($\leq$ 1 Gyr) as a property of star which be same as SD89.
According to the age estimation from observation, which given in the
introduction, we constraint the age of models in the range ~ 0.60
Gyr -- 1.0 Gyr. At last, 12 tracks fall within the observational
error box and are plotted in Fig. 1 (a).

In order to deduce the set of parameters lending to the better
agreement with observations, we perform a $\chi^{2}$ minimization
method described by Eggenberger (\cite{Eggenberger}). We define the
$\chi^{2}$ function as follows:
\begin{equation}
\chi^{2}\equiv\sum\limits^{4}_{i=1}(\frac{C_{i}^{theo}-C_{i}^{obs}}{\sigma})^{2}
\end{equation}
where the vector $C_{i}^{obs}$ contains the observations:
$C_{i}^{obs}$ $\equiv$ ($L/L_{\odot}$, $T_{eff}$, $R/R_{\odot}$,
$[Z/H]_{s}$). The value $C_{i}^{obs}$ and the vector $\sigma$ which
contains the errors on the observations are given in Table 1. From
this method we obtain three models A3, B3 and C3 which are given in
Fig. 1 (b) and Table 2.

\subsubsection{Calibration of models with and without diffusion}

In order to investigate the effect of element diffusion, especially
metal diffusion, basing on the selected models A3, B3 and C3, we
calibrate three groups of models with mass $M/M_{\odot}=0.83$. There
are the different initial chemical compositions among the different
groups. The models at the same group have the same initial chemical
compositions, but they have different types of element diffusion.
The detail description of the models are given in Table 2.

We use the stellar pulsation code of Guenther (\cite{Guenther94}) to
calculate the eigenfrequencies and the large spacings of models
listed in Table 2. For solar-like stars, the eigenfrequencies
$\nu_{n,l}$ of oscillation modes, characterized by the radial order
$n$ at harmonic degree $l$, satisfy the simplified asymptotic
relation (Tassoul \cite{Tassoul}):
\begin{equation}
\nu_{n,l}=\Delta
\nu(n+\frac{l}{2}+\alpha+\frac{1}{4})+\epsilon_{n,l}
\end{equation}
The large spacings is defined by the frequencies of the same
harmonic degree $l$ and the adjacent radial order $n$:
\begin{equation}
\Delta\nu_{n,l}\equiv\nu_{n,l}-\nu_{n-1,l}
\end{equation}
We average the large spacings over the modes $l$ = 0, 1, 2, 3 and
$n$ = 10, 11, ..., 30 to obtain the average large spacings.

From the age of the star in Table 2, we can see that the diffusion
could speed up the evolution of stars. The model with both helium
and metal diffusion has the smallest age of all the models in the
same group. The average large spacings is about 194 $\pm$ 1$\mu Hz$
of models A3, B3 and C3. The pure helium diffusion hardly affect
$\nu_{n,l}$, but the effect is relatively obvious after adding the
metal diffusion. It also seems that the diffusion effect is increase
alone with the increase of the initial heavy element abundance
($Y_{i}$, $Z_{i}$). In the following section, we select models in
group (C) to discuss differences in their internal structures in
detail using the asteroseismic method.

\section{Asteroseismic test helium and metal diffusion effects}

\subsection{``Second differences" tests helium and metal gradients}
   \begin{figure}
\includegraphics[angle=0,totalheight=6.5cm,width=11cm]{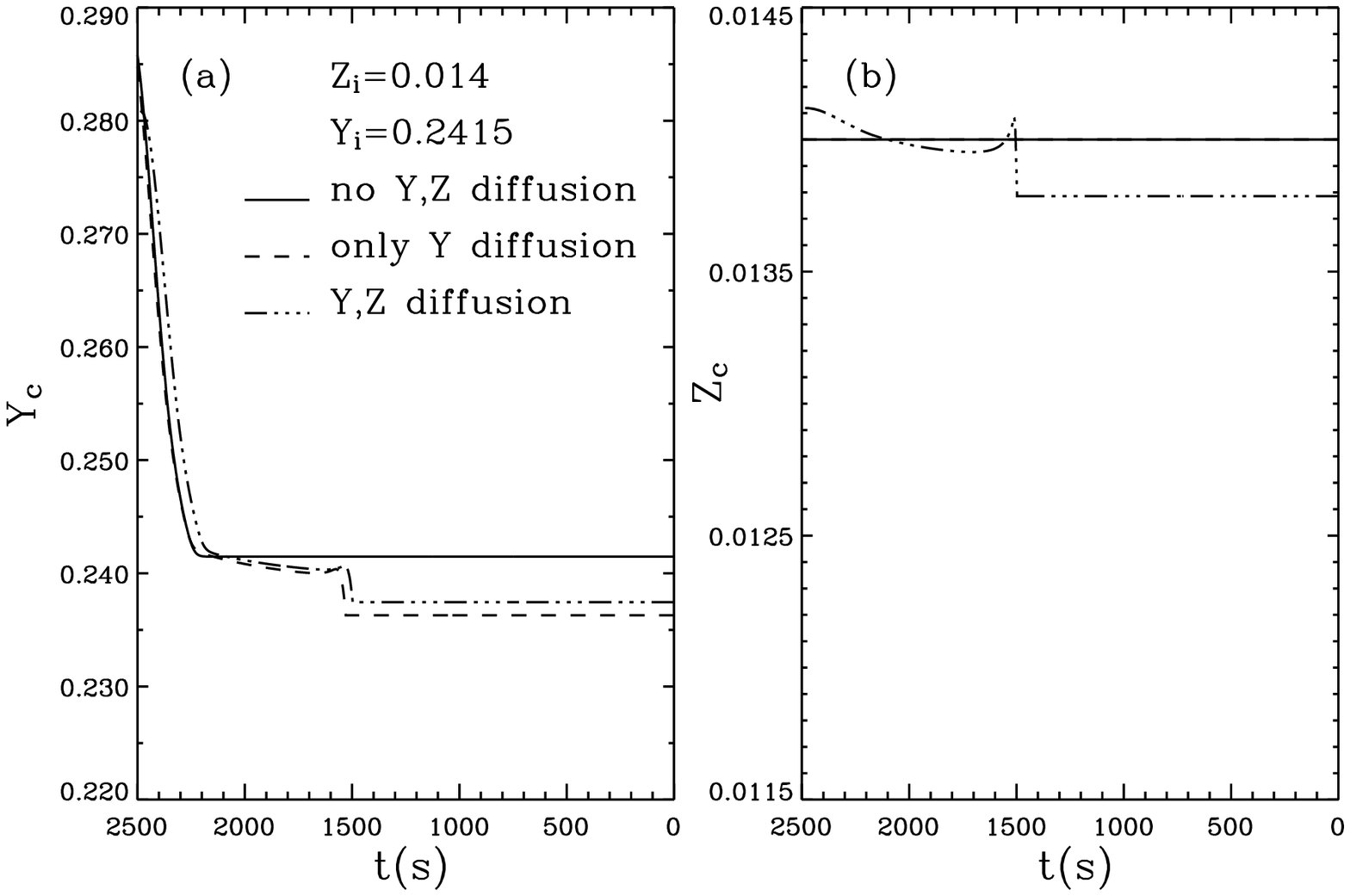}
\vskip 0.15in
\includegraphics[angle=0,totalheight=6.5cm,width=11cm]{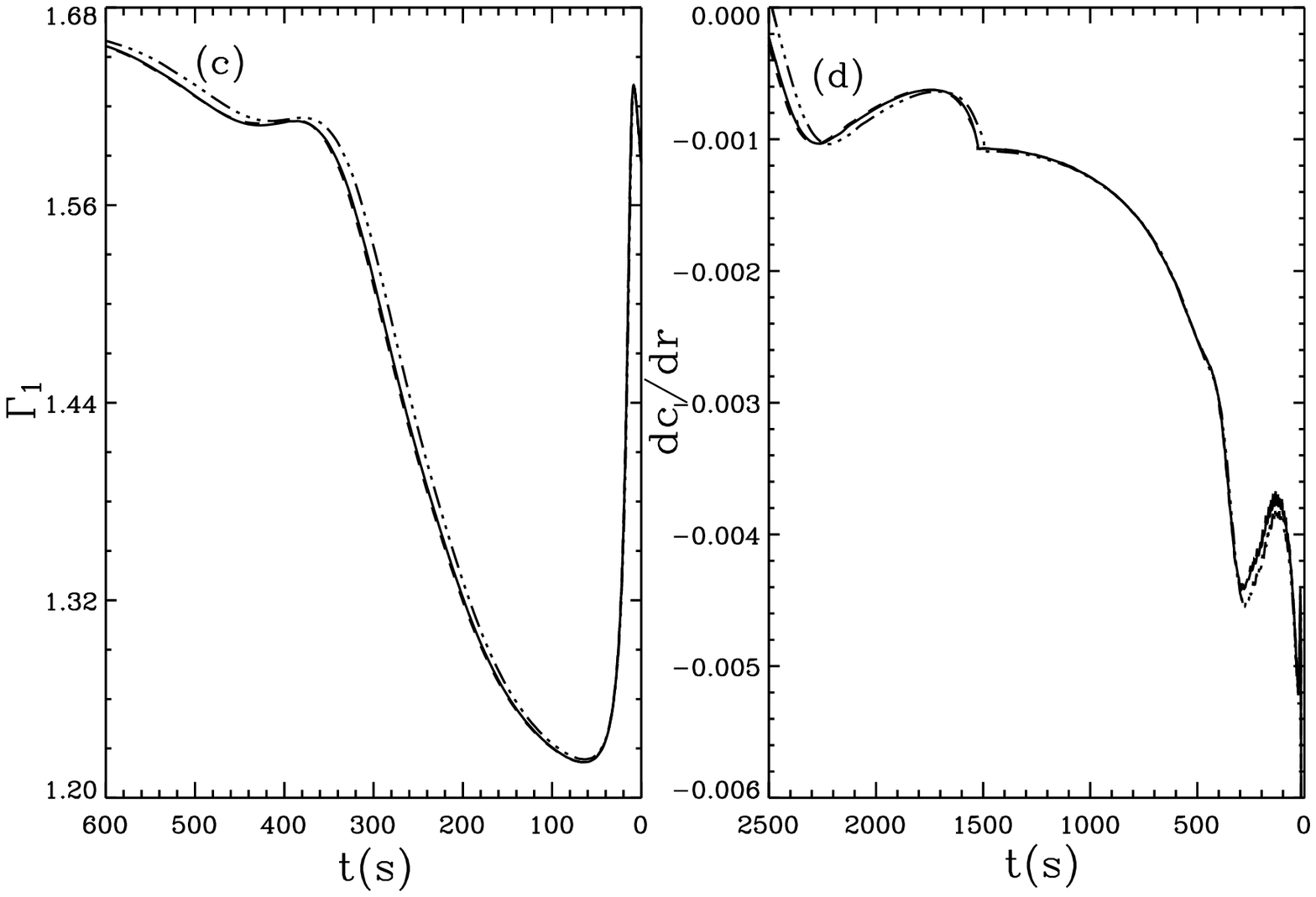}
\vskip 0.15in
\includegraphics[angle=0,totalheight=6.5cm,width=11cm]{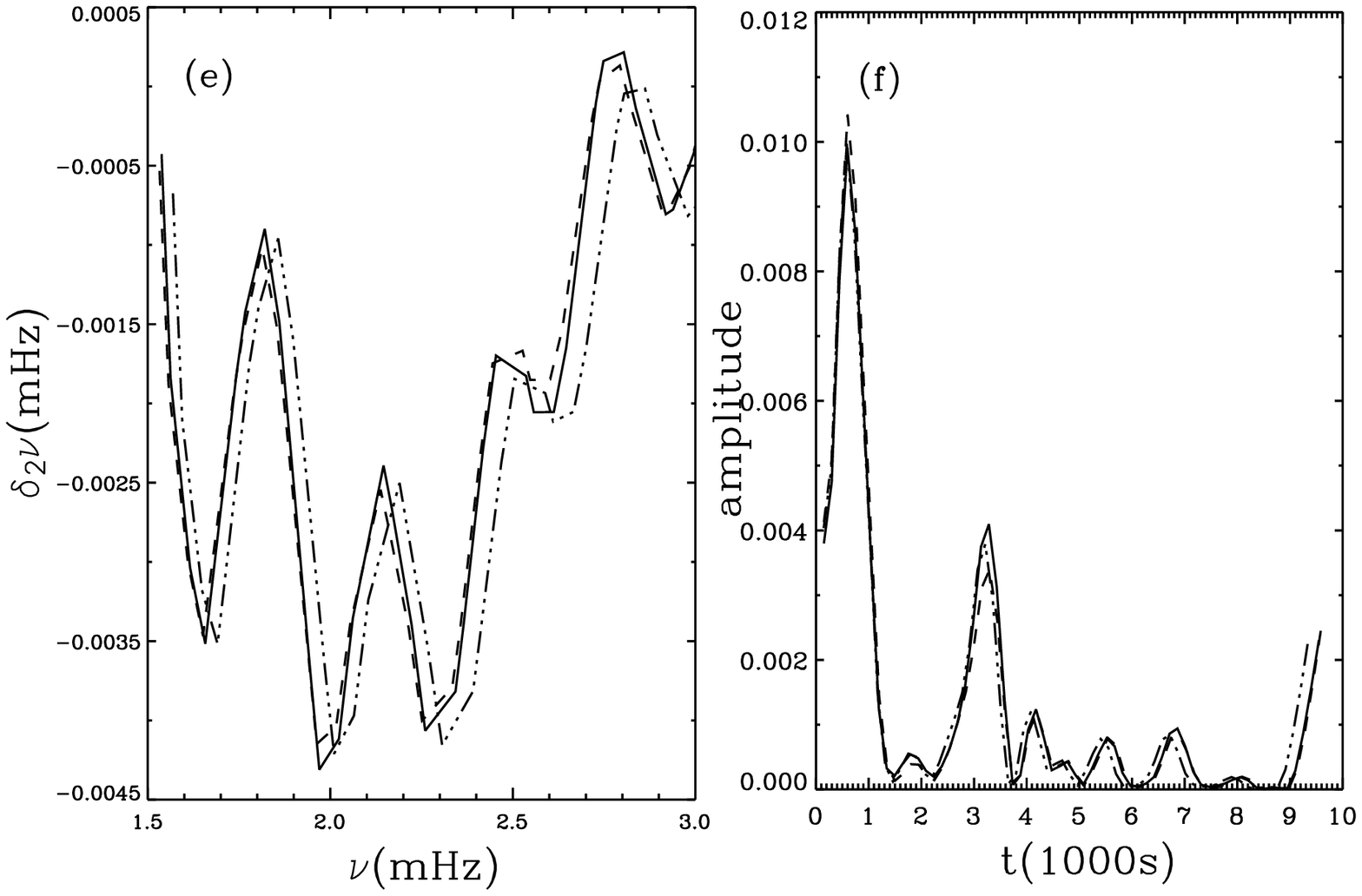}
\vskip 0.15in \caption{The stellar structures of the group (C)
models in Table 2 with $Y_{i}$ = 0.2415, $Z_{i}$ = 0.014. The
detailed notes are labeled in the figure. Solid lines: standard
homogeneous model C1 without diffusion; dashed lines: model C2 with
pure helium diffusion; dash dot lines: model C3 with both helium
diffusion and metal diffusion. The panel (a) helium profiles as a
function of the acoustic depths in the models, i.e. the time for the
acoustic waves to travel from the surface down to the considered
layer; (b) the metal profiles: a clear gradient at the bottom of the
convection zone; (c) the adiabatic exponent $\Gamma_{1}$ profiles
which show a clear feature at the place of the He$_{\rm II}$
ionisation zones; (d) the sound speed gradients: the dips are
clearly visible at the place of the He$_{\rm II}$ ionisation zones
and the bottom of the convection zones; (e) the second differences
of the oscillation frequencies plotted as a function of the
frequencies; (f) the Fourier transform of graph (e).}
\end{figure}
Stellar acoustic $p$-modes with low $l$ degree can propagate deeply
inside the stars. However, as mentioned by Gough (1990), rapid
variations of the sound speed inside a star lead to partial
reflections of the sound waves. A conveniently and easily evaluated
measure of this oscillatory component is the ``second differences"
with respect to the radial order $n$ of the frequencies $\nu_{nl}$
for the same value of the harmonic degree $l$:
\begin{equation}
\Delta_{2}\nu_{nl}\equiv \nu_{n-1,l}-2\nu_{n,l}+\nu_{n+1,l}
\end{equation}
This measure is contaminated less than the first difference
$\Delta_{1}\nu_{nl}$ $\equiv$ $\nu_{n,l}$ - $\nu_{n-1,l}$ by the
smoothly varying components of $\Delta_{1}\nu_{nl}$. This modulation
in the frequency has been extensively studied in the He$_{\rm II}$
ionization zone and the edge between the convective and radiative
zones. In the He$_{\rm II}$ ionization zone there is a rapid
variation of the adiabatic exponent $\Gamma_{1}$ and in the base of
the convection zone there is essentially a discontinuity in the
sound speed gradient, i.e. $dc/dr$, produce a discontinuity in
$\nu_{nl}$. In order to identify the different components which
modulate the oscillations, we have computed the Fourier transform of
the ``second differences". The modulation period of the ``second
differences" is twice the ``acoustic depth" of the region where the
feature occurs:
\begin{equation}
\tau_{s}=\int^{R}_{r_{s}}\frac{dr}{c(r)}
\end{equation}
where $\tau_{s}$ is the time needed for the acoustic waves to travel
between the surface and the considered region (the He$_{\rm II}$
ionization zone or the base of the convection zone), $c(r)$ is the
sound speed at radius $r$, and $r_{s}$ is the radius of the
considered region.

Vauclair \& Th\'{e}ado (\cite{Vauclair}), Th\'{e}ado et al.
(\cite{Theado}) and Castro \& Vauclair (\cite{Castro}) described
asteroseismic signatures of pure helium diffusion in stars between
1.1$M_{\odot}$ and 2.0$M_{\odot}$ in terms of ``second differences".
They have studied the precise signatures on the oscillation
frequencies of helium gradients inside stars. In our work we compute
stars of 0.83$M_{\odot}$ in which not only including helium
diffusion but also containing metal diffusion. The aim of the
present section is to study the precise signatures on the
oscillation frequencies of not only helium gradients but also metal
gradients inside stars and to test the effect of metal diffusion on
the internal structure.

Fig. 2(a)(b) display the helium and metal abundance profiles in the
models without element diffusion (solid line), with pure helium
diffusion (dashed line) and with both helium and metal diffusion
(dash dot line) as a function of the acoustic depth. Due to
diffusion, helium and metal drift inward and form a gradient just
below the convective zone. The acoustic depth of the bottom of the
convection zones $\tau_{cz}$ and the corresponding
$\tau_{cz}/\tau_{0}$ , where $\tau_{0}$ is the total acoustic depth
of the star, are listed in Table 2 about models with different
initial chemical composition and different types of diffusion. Fig.
2(c)(d) give the adiabatic exponent $\Gamma_{1}$ profiles and the
sound speed gradients $dc/dr$ versus the acoustic depth. From the
$dc/dr$ profiles it is clear to see that the dips around 300s and
1500s are caused by the He$_{\rm II}$ ionization zones and the
bottom of the convection zones.

Considering the asymptotic approximation validity, we select modes
of degrees $l$ = 0, 1, 2, 3 and oscillation frequencies between
1500$\mu Hz$ and 3000$\mu Hz$ to compute the ``second differences".
In Fig. 2(e)(f) we present the second differences of our models and
the Fourier transforms of these curves. We have explained that the
modulation period corresponding to the peaks in the Fourier
transforms are twice the ``acoustic depth" of the region where the
feature occurs. By comparing the dips of $dc/dr$ profiles in Fig.
2(d) with Fourier transforms in Fig. 2(f), one can easily recognize
the peaks due to the He$_{\rm II}$ ionization zones and those due to
the bottom of the convective zones in Fig. 2(f).

From the Fourier transforms in Fig. 2(f), we can see that the
amplitudes of the peaks corresponding to the modulation periods due
to the base of the convection zones of models C1, C2, C3 are
different. Mazumdar \& Antia (\cite{Mazumdar}) shown that the
amplitude of the oscillatory signal in the second differences
contains an amplification factor of
$4\sin^{2}(\pi\tau_{s}/\tau_{0})$. When $\tau_{s}/\tau_{0}
> \frac{1}{2}$, the Fourier transforms show the peak amplitudes
decrease for deeper layers (Vauclair \& Th\'{e}ado \cite{Vauclair}).
In terms of $r_{cz}/R$ and $\tau_{cz}/\tau_{0}$ in Table 2, we find
that diffusion could deepen the outer convective zone, especially
including the metal diffusion, the base of the convective zone is
deepest of all in the same group. From the Fourier transforms, we
can see that the peak amplitude at $2\tau_{cz}$ of models with pure
helium diffusion is lower than that without diffusion. In term of
amplification factor of $4\sin^{2}(\pi\tau_{s}/\tau_{0})$, the peak
amplitude of models with both helium and metal diffusion should be
lowest of all, but it is higher than that with pure helium
diffusion. So we think that the peak amplitude is mainly determined
by the depth of the base of the convection zone, meanwhile it is
also sensitive to the element gradient at the bottom of convective
zone. From Fig. 2(b), it is clear to see that the metal gradient
steepen more rapidly than helium gradient. There is a stronger
reflection of the sound waves in the region of the metal gradient,
that is why the model contain metal diffusion have a higher
amplitude of the peak of $\tau_{cz}$.


\subsection{Tests of internal structure}
\begin{figure}
\centering
\includegraphics[angle=0,totalheight=5cm,width=8cm]{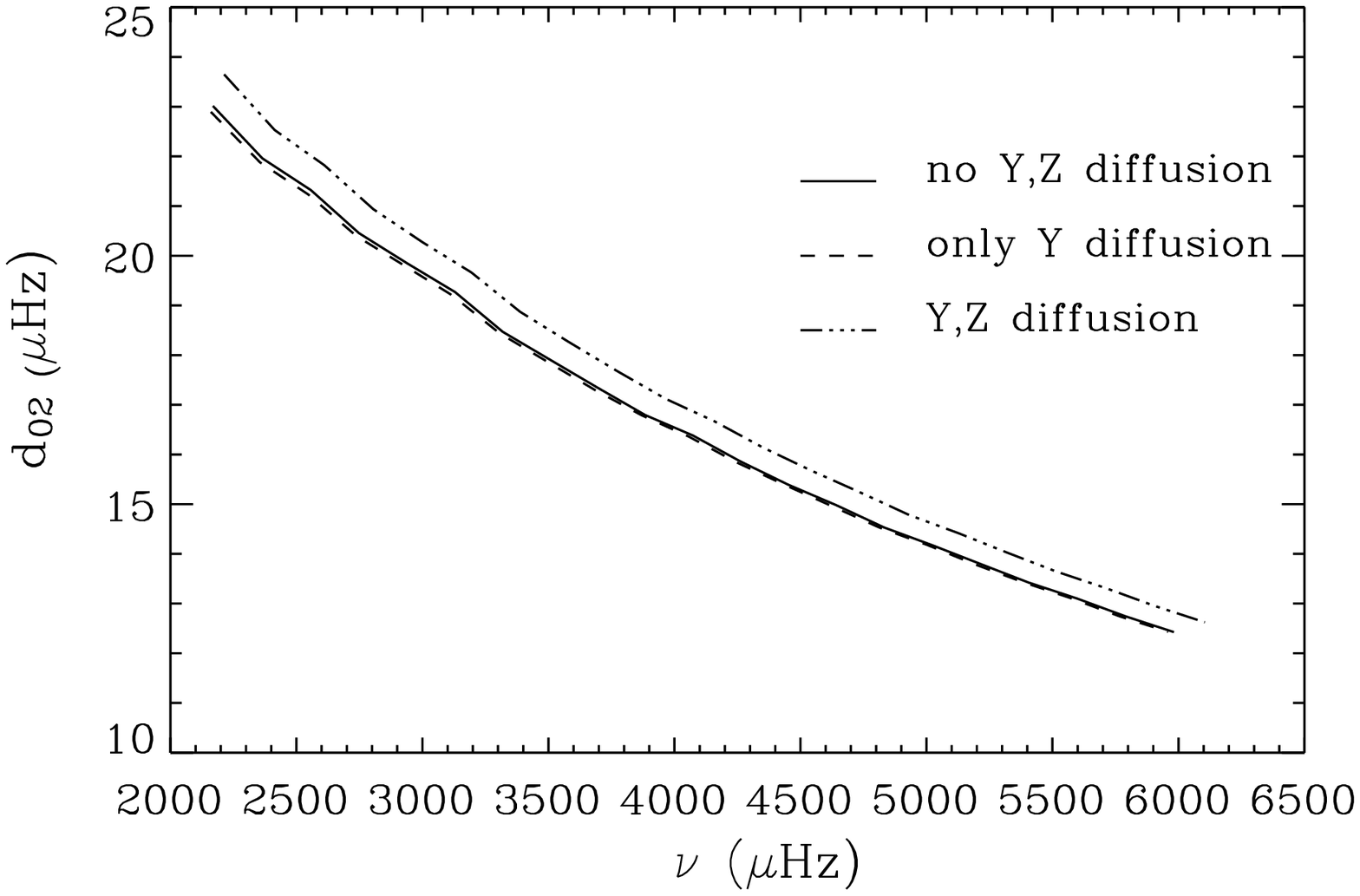}
\vskip 0.30in
\includegraphics[angle=0,totalheight=5cm,width=8cm]{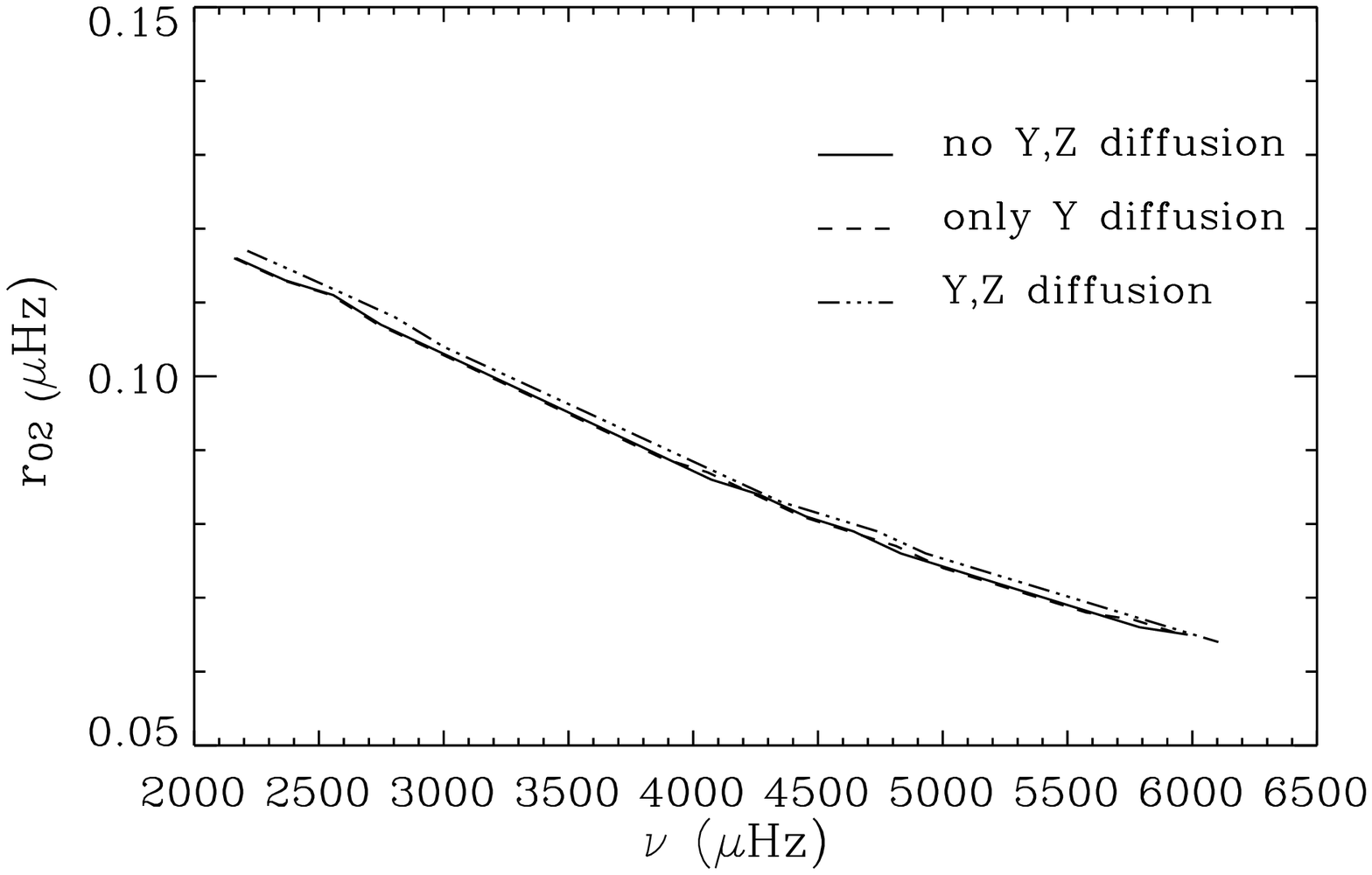}
\vskip 0.20in
 \caption{Upper part: small spacings $d_{02}(n)$, lower
part: frequency separation ratios $r_{02}(n)$ for the models in
group (C). The detailed notes are labeled in the figure.}
\end{figure}

\begin{figure}
\centering
\includegraphics[angle=0,totalheight=4.6cm,width=8cm]{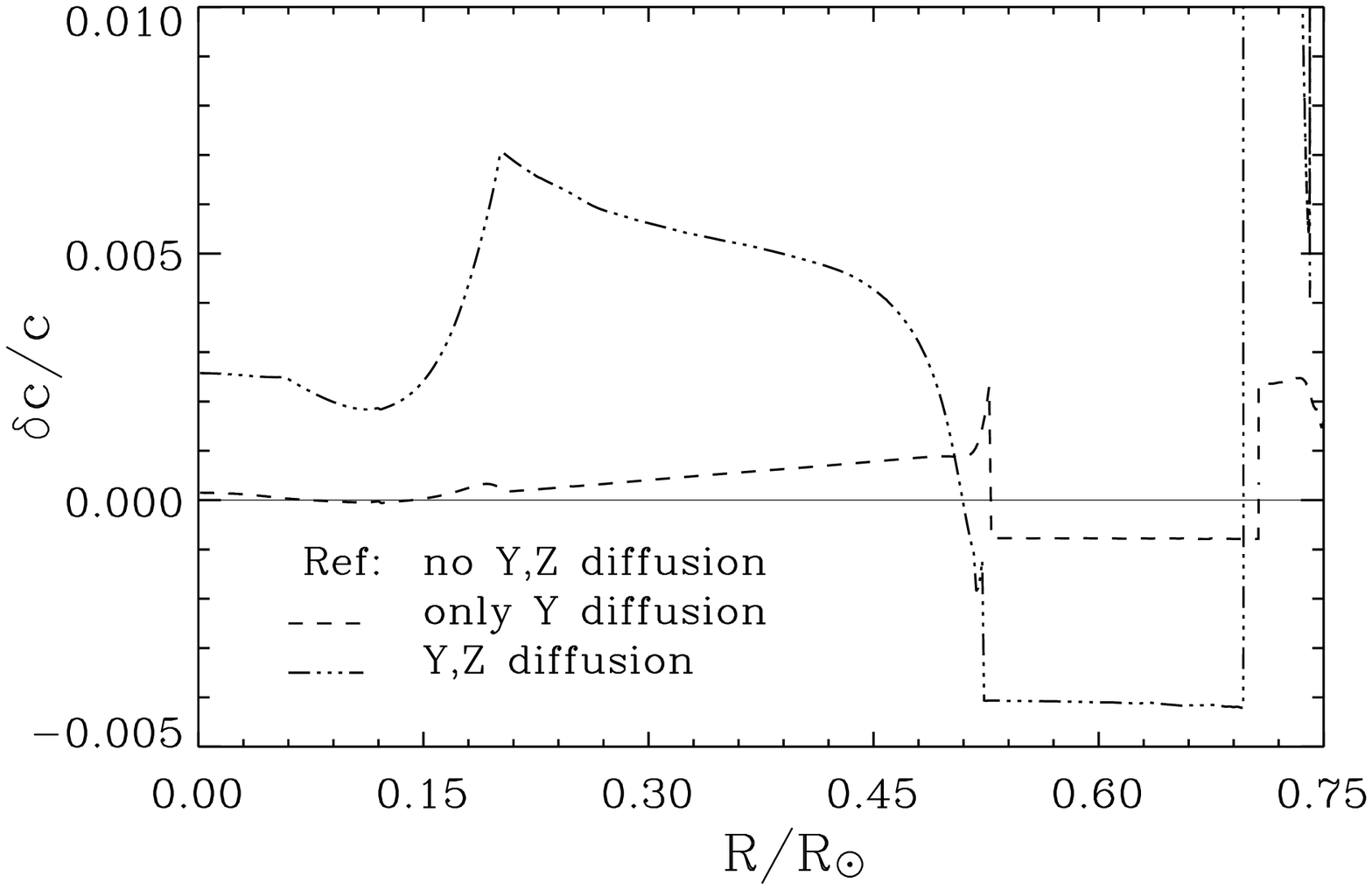}
\vskip 0.20in
\includegraphics[angle=0,totalheight=4.6cm,width=8cm]{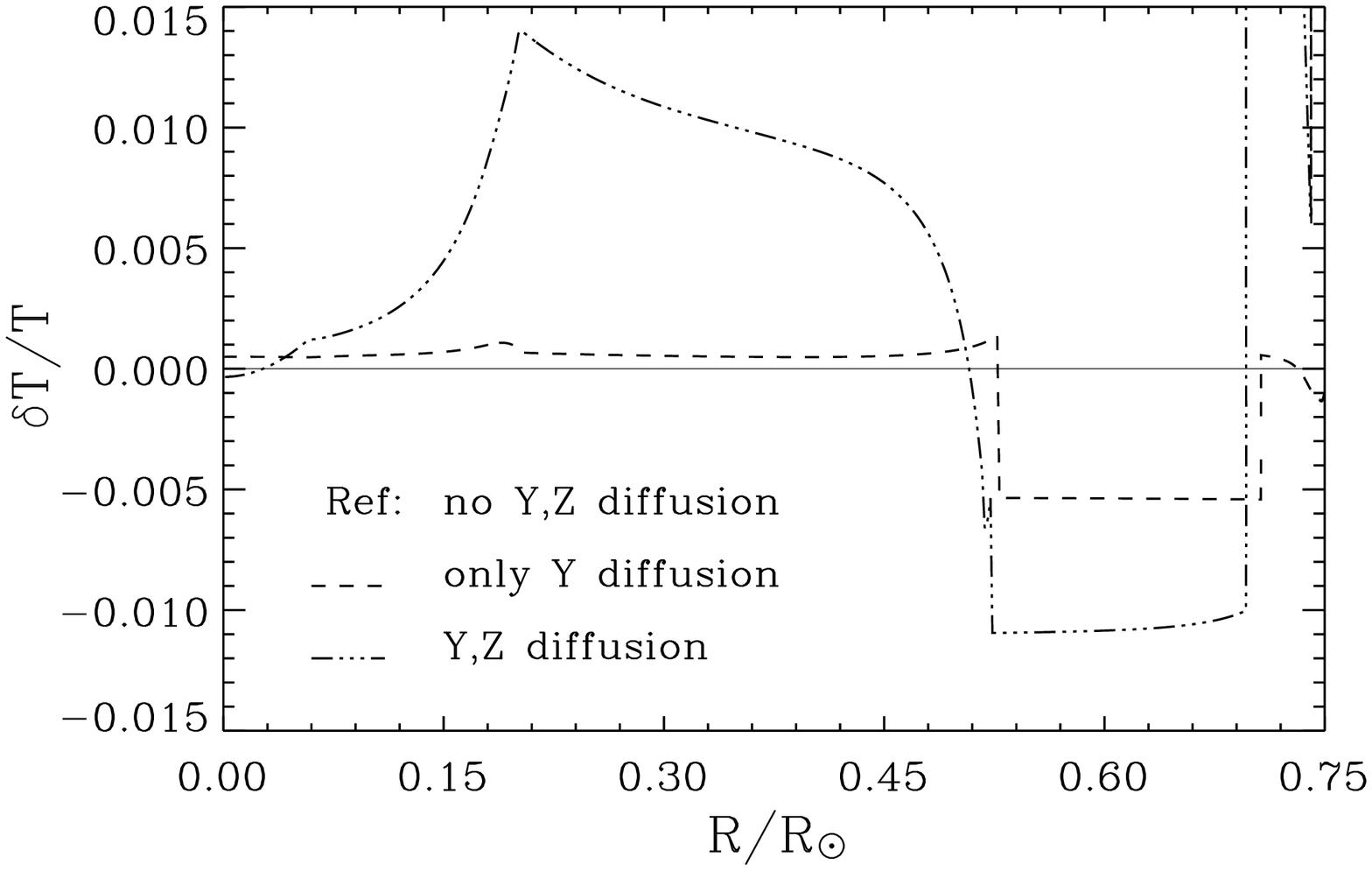}
\vskip 0.20in
\includegraphics[angle=0,totalheight=4.6cm,width=8cm]{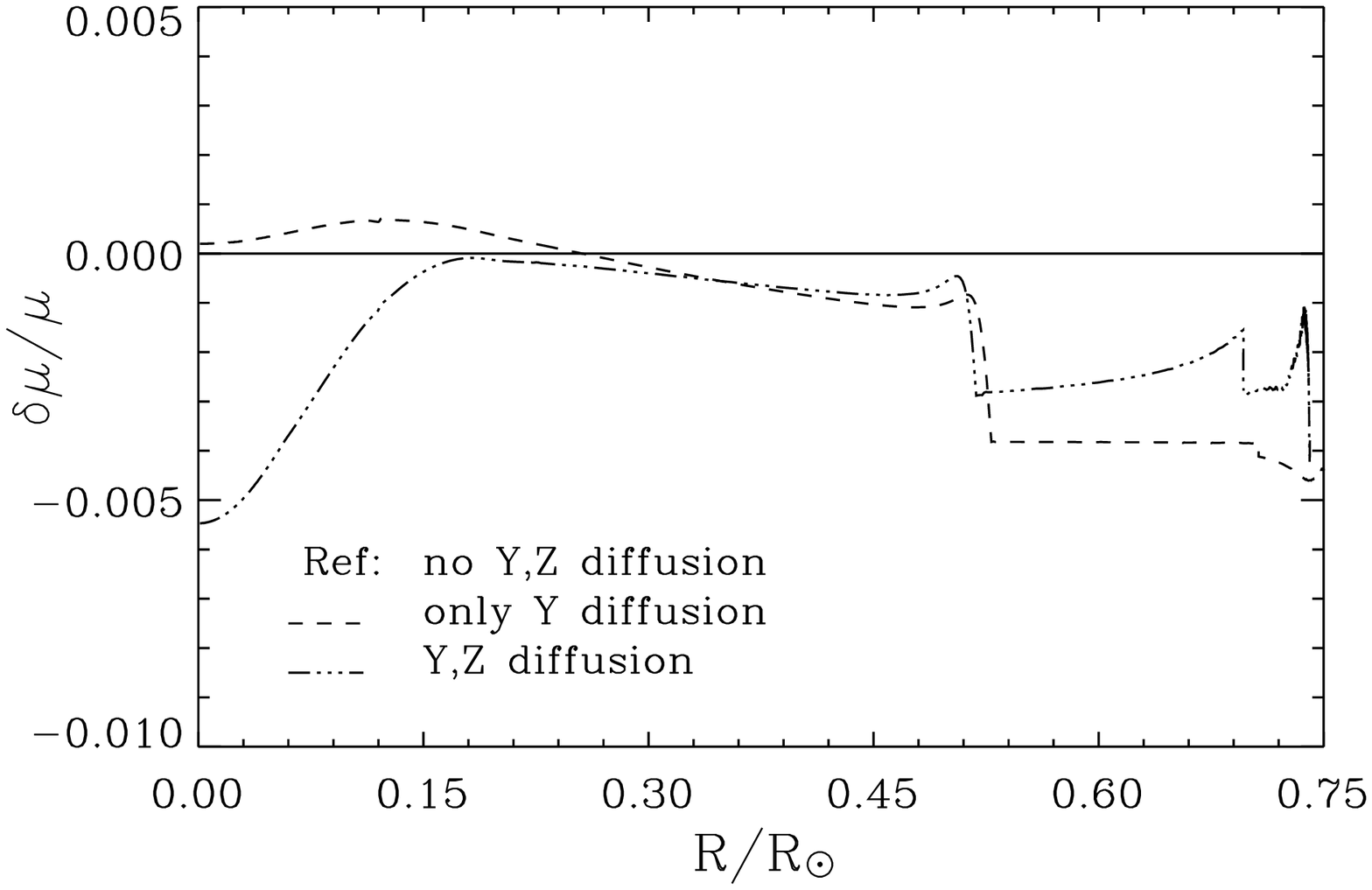}
\vskip 0.20in
 \caption{From up to down, each panel of figure give
the relative sound-speed difference $\delta c/c$, the temperature
difference $\delta T/T$ and the mean molecular weight $\delta
\mu/\mu$ respectively for the models in group (C). All differences
are with respect to the model C1 which without any diffusion. The
dashed lines render the differences between the model with pure $He$
diffusion and the model without any diffusion (e.g.  $\delta
\mu/\mu$ = ($\mu_{Y.Dif}$ - $\mu_{None. Dif}$) / $\mu_{None. Dif}$).
The dash dot lines render the differences between the model with
$He$ \& $Z$ diffusion and the model without any diffusion (e.g.
$\delta \mu/\mu$ = ($\mu_{Y \& Z.Dif}$ - $\mu_{None. Dif}$) /
$\mu_{None. Dif}$). }
\end{figure}
In this section, we compute the so-called small frequency spacings
and frequency separation ratios of the low-$l$ $p$-modes to test the
internal structure of models in which considered different types of
diffusion. The small frequency spacings is given by the combination:
\begin{equation}
d_{ll+2}(n)=\nu_{n,l}-\nu_{n-1,l+2}
\end{equation}
Using the asymptotic theory of $p$ modes it can be shown that
(Christensen-Dalsgaard \& Berthomieu
\cite{Christensen-Dalsgaard91}; Basu et al. \cite{Basu07}):
\begin{equation}
d_{ll+2}(n)\simeq
-(4l+6)\frac{\Delta\nu_{nl}}{4\pi^{2}\nu_{n,l}}\int^{R}_{0}\frac{dc}{dr}\frac{dr}{r}
\end{equation}
where $R$ is the stellar radius and $\Delta\nu_{nl}$ is the large
spacings. In the core the gradient of the sound speed is large and
$r$ is small, the integral in equation (14) is dominated by
conditions in the core. So the small spacings usually test the
stellar core. But the small spacings are also slightly affected by
near-surface effects. In order to reduce the effect of near surface
uncertainties, the frequency separation ratios are used (Roxburgh \&
Vorontsov \cite{Roxburgh}):
\begin{equation}
r_{02}(n)=\frac{d_{02}(n)}{\Delta_{1}(n)},\hspace{0.5cm}r_{13}(n)=\frac{d_{13}(n)}{\Delta_{0}(n+1)}
\end{equation}

Each panel of Fig. 3 show small spacings $d_{02}$ and frequency
separation ratios $r_{02}$ for different models in group (C) with
different types of diffusion which are labeled in the figure. From
the small spacings and frequency separation ratios we can see that
the pure helium diffusion hardly alter the internal structure in
these young low--mass models. By comparing the models included metal
diffusion with the models no element diffusion, the differences of
their internal structures are relatively obvious.

In order to explain the changes of the internal structure from the
asteroseismic test, we compare the internal physical parameters of
models with and without diffusion. It is known that the sound speed
depends on both the mean molecular weight and the temperature (Basu
et al. \cite{Basu07}; Bi et al. \cite{Bi} )
\begin{equation}
c^{2}\simeq\frac{K_{B}T}{\mu m_{u}}\propto\frac{T}{\mu}
\end{equation}
In Fig. 4, we give the $\delta c/c$, $\delta T/T$, and $\delta
\mu/\mu$ variations as function of radius.

From Fig. 4, it is clear to see that the effect of the pure helium
diffusion on the sound speed $c$, temperature $T$ and the mean
molecular weight $\mu$ is small. The helium diffusion hardly alter
the $c$, $T$ and $\mu$ in the radiation interior and the effect
mainly concentrate on the convection zone. Due to diffusion, helium
drift inward just below the convection zone. It induce $\mu$ is
smaller than that without diffusion in the envelope. On the other
hand, the increase of the opacity in the envelope due to the
settling of $He$ leads to the lower temperature than that without
diffusion. However, the helium diffusion decrease both the $T$ and
$\mu$, the sound speed $c$ slightly in the convective envelope. So
the small spacings and separation ratios are similar with and
without helium diffusion.

In Fig. 4 the dash dot lines describe the differences of $c$, $T$
and $\mu$ between the model with both helium and metal diffusion and
the model without any diffusion. From Fig. 4, we can see that the
effect of metal diffusion on the $c$, $T$ and $\mu$ not only in the
convective envelope but also in the radiation interior. Due to
diffusion, helium and metal drift inward just below the convection
zone and decrease $\mu$ in the outer envelope. At the same time, the
depletion of $Z$ increase the opacity in the envelope, otherwise the
concentration decrease the opacity in the radiation interior. It
leads to lower temperature in the envelope and higher value in the
interior. From Fig. 4, we can see that the temperature is more
sensitive to the metal diffusion than other physical parameters. The
$\delta c/c$ is mainly dependent on the variation of $\delta T/T$.
Except the convective envelope, the sound speed $c$ is much larger
in the interior of the model. From  the expression of the sound
speed gradient in terms of the equation (16):
\begin{equation}
\frac{dc}{dr} \propto  \frac{d(\frac{T}{\mu})^{1/2}}{dr}\\
=-\frac{1}{2}(\frac{T}{\mu^{3}})^{\frac{1}{2}}\frac{d\mu}{dr}+\frac{1}{2}\frac{1}{(T
\mu)^{\frac{1}{2}}}\frac{dT}{dr}
\end{equation}
 we find that the temperature $T$ is the dominate factor in the
 sound speed gradient in causing the differences between the models
 with and without metal diffusion. Due to $dT/dr$ is negative,
 $|dc/dr|$ is larger than that of model without metal
 diffusion in the radiation interior. So from equation (14), the
 small spacings are larger than other models. That are explained the
 variation of
 the internal structure including metal diffusion.

\section{Conclusions}

Taking into account the effects of helium and metal diffusion on
stellar evolution, we have presented a detailed modeling of the MOST
target: the star $\epsilon$ Eridani by fitting the available
observational constraints: effective temperature, luminosity, high
precision interferometric observational radius and metallicity. We
also consider the age ($\leq$ 1 Gyr) as a property of star to
constraint the models, which are the same as SD89.

According to our results, the location in the H-R diagram indicates
that the star $\epsilon$ Eridani has evolved little since its
arrival on the zero-age main sequence. Taking into account the
helium and metal diffusion, we find that the mean large spacings
$<\Delta\nu>$, which are averaged over $l$ = 0, 1, 2, 3 and $n$ =
10, 11, 12, ..., 30, is about 194 $\pm$ 1 $\mu Hz$.

In order to test the effect of helium and metal diffusion, we
construct three groups of models with the same mass but different
$Y_{i}$ and $Z_{i}$. Then, for the given chemical composition we
calibrate three models with the different types of diffusion which
are listed in Table 2. In our work we fixed the mixing length
parameter $\alpha$ = 1.70 and neglected convective overshoot. The
results show that the age of diffused models is about 1 Gyr, and be
younger than that determined by the models without diffusion.

Due to the effect of diffusion, helium and metal fall below the
outer convective zone and form a gradient. We use the ``second
differences" to test the depth of the convective zones. When metal
diffusion is considered, the convective zone deepens more quickly
and the metal gradient steepens more rapidly than the helium
gradient. That induce a stronger reflection of the sound waves
just below the convection zone, which explains the higher
amplitude of the peak related to the base of the convective zone
than that of pure helium diffusion.

The small spacings and frequency separation ratios have been used to
test the differences of internal structure in models with and
without diffusion. According to our results, the pure helium
diffusion hardly alter the internal structure in the young low-mass
models. Taking into account the metal diffusion, it mainly induce a
much higher temperature in the radiation interior, thereby the
higher sound speed in the interior of the model. Correspondingly,
the frequencies and large spacings are larger obviously than that of
models without metal diffusion. The variation of internal structures
between models with and without metal diffusion is increase
obviously alone with the increase of initial metallicity. So we
conclude that it is necessary to consider the metal diffusion in the
young main-sequence low-mass star.

\begin{acknowledgements}
 We are grateful to anonymous referee for his/her constructive
suggestions and valuable remarks to improve the manuscript. This
work was supported by The Ministry of Science and Technology
 of the People's republic of China through grant 2007CB815406,
 and by NSFC grants 10173021, 10433030, 10773003, and 10778601.
\end{acknowledgements}

\end{document}